\documentclass[sigconf]{acmart}

\usepackage[linesnumbered,ruled,vlined]{algorithm2e}
\usepackage{subcaption}
\usepackage{float}
\usepackage{multirow}
\usepackage{booktabs}
\usepackage{bm}

\AtBeginDocument{%
  \providecommand\BibTeX{{%
    \normalfont B\kern-0.5em{\scshape i\kern-0.25em b}\kern-0.8em\TeX}}}

\setcopyright{acmcopyright}
\copyrightyear{2018}
\acmYear{2018}
\acmDOI{10.1145/1122445.1122456}

\acmConference[Woodstock '18]{Woodstock '18: ACM Symposium on Neural
  Gaze Detection}{June 03--05, 2018}{Woodstock, NY}
\acmBooktitle{Woodstock '18: ACM Symposium on Neural Gaze Detection,
  June 03--05, 2018, Woodstock, NY}
\acmPrice{15.00}
\acmISBN{978-1-4503-XXXX-X/18/06}

\begin{document}

\title{EPIC30M: An Epidemics Corpus Of Over 30 Million Relevant Tweets}


\author{Junhua Liu}
\affiliation{%
  \institution{Singapore University of Technology and Design}}
\additionalaffiliation{
  \institution{Forth AI, j@forth.ai}}
\email{junhua_liu@mymail.sutd.edu.sg}

\author{Trisha Singhal}
\affiliation{%
  \institution{Singapore University of Technology and Design}}
\email{trisha_singhal@sutd.edu.sg}

\author{Lucienne T.M. Blessing}
\affiliation{%
  \institution{Singapore University of Technology and Design}}
\email{lucienne_blessing@sutd.edu.sg}

\author{Kristin L. Wood}
\affiliation{%
  \institution{Singapore University of Technology and Design}}
\additionalaffiliation{
  \institution{University of Colorado Denver, kristin.wood@ucdenver.edu}}
\email{kristinwood@sutd.edu.sg}

\author{Kwan Hui Lim}
\affiliation{%
  \institution{Singapore University of Technology and Design}}
\email{kwanhui_lim@sutd.edu.sg}
\authornote{Corresponding author.}


\renewcommand{\shortauthors}{Liu et al.}

\begin{abstract}

Since the start of COVID-19, several relevant corpora from various sources are presented in the literature that contain millions of data points. While these corpora are valuable in supporting many analyses on this specific pandemic, researchers require additional benchmark corpora that contain other epidemics to facilitate cross-epidemic pattern recognition and trend analysis tasks. During our other efforts on COVID-19 related work, we discover very little disease related corpora in the literature that are sizable and rich enough to support such cross-epidemic analysis tasks. In this paper, we present EPIC30M, a large-scale epidemic corpus that contains 30 millions micro-blog posts, i.e., tweets crawled from Twitter, from year 2006 to 2020. EPIC30M contains a subset of 26.2 millions tweets related to three general diseases, namely Ebola, Cholera and Swine Flu, and another subset of 4.7 millions tweets of six global epidemic outbreaks, including 2009 H1N1 Swine Flu, 2010 Haiti Cholera, 2012 Middle-East Respiratory Syndrome (MERS), 2013 West African Ebola, 2016 Yemen Cholera and 2018 Kivu Ebola. Furthermore, we explore and discuss the properties of the corpus with statistics of key terms and hashtags and trends analysis for each subset. Finally, we demonstrate the value and impact that EPIC30M could create through a discussion of multiple use cases of cross-epidemic research topics that attract growing interest in recent years. These use cases span multiple research areas, such as epidemiological modeling, pattern recognition, natural language understanding and economical modeling. \footnote{The corpus is available at https://www.github.com/junhua/epic.}

\end{abstract}





\maketitle


\begin{table*}[t]
\centering
\setlength{\tabcolsep}{15pt}
\resizebox{\textwidth}{!}{%
\begin{tabular}{lcccc}
\hline
Epidemics            &       \multicolumn{1}{c}{Queries}     & Earliest tweet & Latest tweet             & No. of tweets        \\ \hline

\multicolumn{4}{l}{\textit{\textbf{Outbreak}}} \\
2009 H1N1 Swine Flu          &   h1n1                & 05-01-2009        & 19-06-2020        & 2,803,941   \\
2010 Haiti Cholera           &   haiti cholera       & 13-01-2010        & 31-05-2020        & 359,122      \\
2012 Middle East Respiratory Syndrome &   MERS-CoV, \#mers    & 01-09-2012        & 31-05-2020        & 265,119      \\
2014 West Africa Ebola       &   africa ebola        & 01-12-2013        & 19-06-2020        & 1,191,516     \\
2016 Yemen Cholera           &   yemen cholera       & 01-01-2017        & 15-05-2020        &  102,900  \\
2018 Kivu Ebola              &   kivu ebola          & 03-03-2018        & 31-05-2020        & 12,063       \\ \hline

\multicolumn{4}{l}{\textit{\textbf{General}}} \\
Cholera             &       cholera       & 19-01-2007                & 19-06-2020      & 2,321,903         \\
Ebola               &       ebola         & 25-12-2006                & 19-06-2020      & 20,178,969        \\
Swine Flu           &       swine flu     &  31-10-2007               & 19-06-2020      & 3,775,217    \\ \hline

\end{tabular}}
\caption{Statistics of EPIC30M with the \textit{Outbreak} and \textit{General} subsets\protect\footnotemark}
\label{tab:search}
\end{table*}

\section{Introduction}

The Coronavirus disease (COVID-19) has spread around the globe since the beginning of the year 2020, affecting around 200 countries and everyone's life. To date, the highly contagious disease has caused over 6.6 million confirmed and suspected cases and 389 thousand deaths. In time of crisis caused by epidemics, we realize the necessity of rigorous arrangements, quick responses, credible and updated information during the premature phases of such epidemics~\cite{world2018managing}. 

Social media platforms, such as Twitter, play an important role in informing the latest epidemic status, via the announcements of public policies in a timely manner. Facilitating the posting of over half a billion tweets daily~\cite{liu2020crisisbert}, Twitter emerges as a hub for information exchange among individuals, companies, and governments, especially in time of epidemics where economies are placed in a hibernation mode, and citizens are kept isolated at home. Such platforms help tremendously to raise situational awareness and provide actionable information~\cite{imran2013extracting}.

Recently, numerous COVID-19 related corpora from various sources are presented that contain millions of data points~\cite{chen2020covid, lopez2020understanding}. While these corpora are valuable in supporting many analyses on this specific pandemic, researchers require additional benchmark corpora that contain other epidemics to facilitate cross-epidemic pattern recognition and trend analysis tasks. During our other efforts on COVID-19 related work, we discovered very little disease related corpora in the literature that are sizable and rich enough to support such cross-epidemic analysis tasks.

In this paper, we present EPIC30M, a large-scale epidemic corpus that contains 11.8 millions micro-blog posts, i.e., tweets crawled from Twitter, from year 2006 to 2020. EPIC30M contains a subset of 3.9 millions tweets related to three general diseases, namely Ebola, Cholera and Swine Flu, and another subset of 7.9 millions tweets of six global epidemic outbreaks, including 2009 H1N1 Swine Flu, 2010 Haiti Cholera, 2012 Middle-East Respiratory Syndrome (MERS), 2013 West African Ebola, 2016 Yemen Cholera, and 2018 Kivu Ebola.

We conduct several exploratory analyses to study the properties of the corpus, such as word cloud visualization and time series trend analysis. Several interesting findings are discovered through these analyses. For instance, we find that a large quantity of topics are related to specific locations; cross-epidemic topics, i.e. one that involves more than one epidemic-related hashtag, appear frequently in several classes; and several hashtags related to non-epidemic events, such as warfare, have relatively high ranks in the list. Furthermore, a time-series analysis also suggests that some of the epidemics, i.e. \textit{2010 Haiti Cholera} and \textit{2018 Kivu Ebola}, show a surge in tweets before the respective start dates of the outbreaks, which signifies the importance of leveraging social media to conduct early signal detection. We also observe that an epidemic outbreak not only leads to rapid discussion of its own, but also triggers exchanges about other diseases. 

EPIC30M fills the gap in the literature where very little epidemic-related corpora are either unavailable or not sizable enough to support cross-epidemic analysis tasks. Through discussing various potential use cases, we anticipate that EPIC30M brings great value and impact to various fast growing computer science communities, especially in natural language processing, data science and computation social science. We also foresee that EPIC30M is able to contribute partially to cross-disciplinary research topics, such as economic modeling and humanity studies. While EPIC30M includes tweets posted throughout the cause of each outbreak available in the corpora, we expect that EPIC30M may serve as a timeless cross-epidemic benchmark.

\begin{figure*}[t]
\centering
\begin{subfigure}{\textwidth}
  \includegraphics[width=\textwidth]{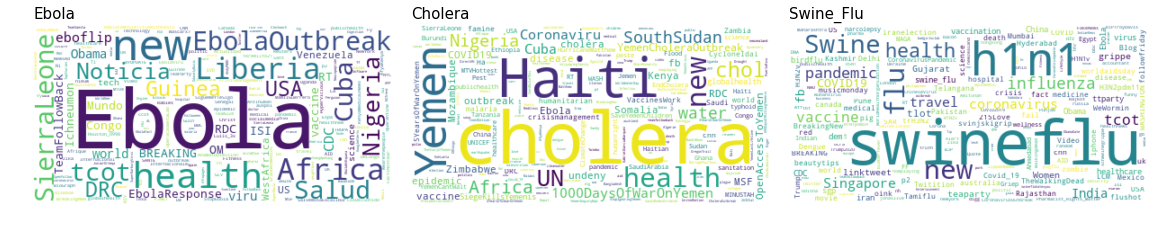}
  \caption{} \label{fig:wc-general}
  \vspace{-.9mm}
  \includegraphics[width=\textwidth]{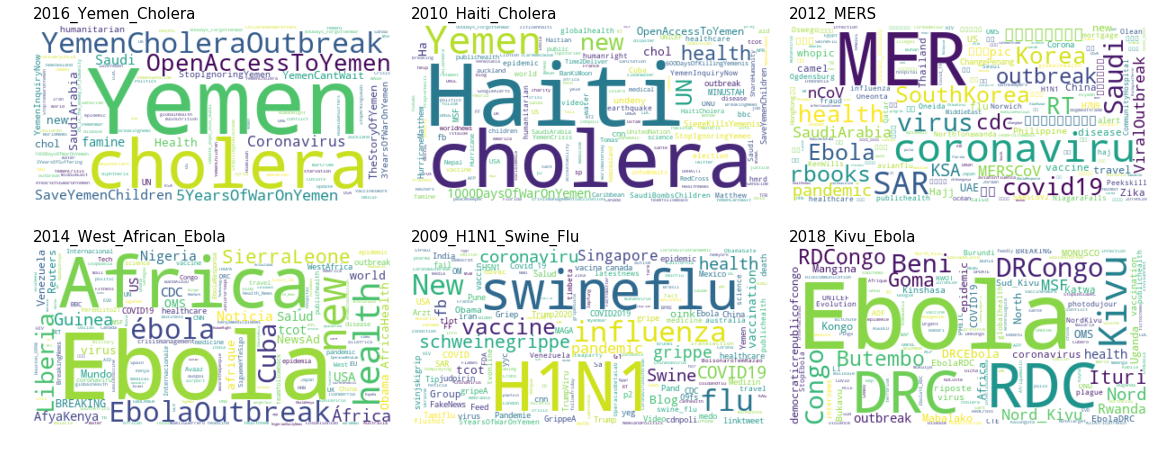}
  \caption{} \label{fig:wc-outbreak}
  \vspace{-1.1mm}
\end{subfigure}
\caption{Hashtags analysis with word clouds~\cite{camburncomputer}. Each word cloud contains the top 100 hashtags in their respective class where the sizes represent the frequency of the hash tags. (a) Three general epidemic classes. (b) Six epidemic outbreak classes.}
\label{fig:worldclouds}
\end{figure*}

\footnotetext{As of 20 Jun 2020}

\section{Related Work}

In this section, we discuss the existing Twitter corpora for several domains, such as COVID-19, disasters, and others. These corpora attract a large quantity of interests and enable a large amount of research works in their respective domains, which we believe EPIC30M generalizes to a similar level of impact in the epidemic domain.

\textbf{Corpora of COVID-19}.
Recently, the COVID-19 pandemic spread across the globe and generated enormous economical and social impact. Throughout the pandemic, numerous related corpora have been released. For instance, \citet{chen2020covid} released a multi-lingual corpus that consists of 50 million tweets that include tweet IDs and their timestamps, across over 10 languages. Similarly, \citet{banda2020large} presented a large-scale COVID-19 chatter corpus that consists of over 152M tweets with retweets and another version of 30 million tweets without retweets.

\textbf{English corpora of disasters}.
There are several disaster-related corpora presented in the literature that are utilized for multiple works. CrisisLex~\cite{olteanu2014crisislex} consists of 60 thousand tweets that are related to six natural disaster events, queried based on relevant keywords and locations during the crisis periods. The tweets are labelled as \textit{relavant} or \textit{not-relevant} through crowdsourcing. \citet{olteanu2015expect} conducts a comprehensive study of tweets to analyze 26 crisis events from 2012 to 2013. The paper analyzes about 25k tweets based on crisis and content dimensions, which include hazard type (\textit{natural} or \textit{human-induced}), temporal development (\textit{instantaneous} or \textit{progressive}), and geographic speed (\textit{focalized} or \textit{diffused}). The content dimensions are represented by several features such as informativeness, types and sources. \citet{imran2016twitter} releases a collection of over 52 million tweets, out of which 50 thousand come with human-annotated tweets that are related to 19 natural crisis events. The work also presents pre-trained \textit{Word2Vec} embeddings with a set of Out-Of-Vocabulary (OOV) words and their normalizations, contributing in spreading situational awareness and increasing response time for humanitarian efforts during crisis. ~\citet{phillips2017hurricane} releases a set of 7 million tweets related to Hurricane Harvey. \citet{littman2018charlottesvill} publishes a corpus containing tweet IDs of over 35 million tweets related to Hurricane Irma and Harvey. 

\textbf{Non-English corpora of disasters}.
Numerous non-english crisis corpora are also found in the literature. For instance, \citet{cresci2015linguistically} released a corpus of 5.6 thousand Italian tweets from 2009 to 2014 during four different disasters. The features include informativeness (\textit{damage} or, \textit{no damage}) and relevance (\textit{relevant} or \textit{not relevant}). Similarly, \citet{alharbi2019crisis} compiled a set of 4 thousand Arabic tweets, manually labelled on the relatedness and information-type for four high risk flood events in 2018.  \citet{alam2018crisismmd} released a Twitter corpora composed of manually-annotated 16 thousand tweets and 18 thousand images collected during seven natural disasters (earthquakes, hurricanes, wildfires, and floods) that occurred in 2017. The features of the datasets include Informativeness, Humanitarian categories, and Damage severity categories. 

\textbf{Other Twitter Corpora}.
Apart from crisis-related corpora, several Twitter datasets are used for analysis related to politics, news, abusive behaviour and misinformation, Trolls, movie ratings, weather forecasting, etc. For instance, \citet{fraisier2018elysee2017fr} proposes a large and complex dataset with over 22 thousand operative Twitter profiles during the 2017 French presidential campaign with their corresponding tweets, tweet IDs, retweets, and mentions. The data was annotated manually based on their political party affiliation, their nature, and gender. We also find Twitter Corpora that are related to other domains, such as politics~\cite{wrubel2019congress, brigadir2015analyzing}, cyberbullying~\cite{founta2018large}, and misinformation~\cite{hui2018hoaxy, zubiaga2016analysing, roeder2018we}.

\section{Data Collection}

This section describes the data collection process for crawling EPIC30M.
 
\textbf{Epidemic Outbreaks}.
EPIC30M includes six epidemic outbreaks in the 21st century, recorded by World Health Organization~\footnote{https://www.who.int/emergencies/diseases/en/} and happened after the founding of Twitter in 2006. These outbreaks include the 2009 H1N1 Swine Flu, the 2010 Haiti Cholera, the 2012 Middle East Respiratory Syndrome (MERS), the 2014 West Africa Ebola, the 2016 Yemen Cholera, and the 2018 Kivu Ebola, as listed in Table~\ref{tab:search}. We intentionally exclude the recent COVID-19 pandemic outbreak to avoid producing redundant work, as there are already numerous COVID-19 datasets released by different parties with multi-million data points. 

\textbf{Search Queries}.
For each outbreak, we initialize with a large collection of keywords used as the search queries, with the hypothesis to retrieve most relevant tweets from Twitter. We use a combination of keywords for each outbreak, as listed on table~\ref{tab:search}, to fetch the related tweets. Two types of keywords used, namely (a) general disease-related terms, such as \textit{ebola}, \textit{cholera} and \textit{swine flu}; and (b) specific outbreak-related terms with a combination of location and disease, such as \textit{africa ebola} and \textit{yemen cholera}. 

\textbf{General Epidemics}. Besides the outbreaks set, we extend EPIC30M by including a subset of three general diseases, namely \textit{Cholera}, \textit{Ebola} and \textit{Swine Flu}. The tweets related to these diseases are crawled since the respective first occurrence until 15$^{th}$ May 2020. We expect the \textit{general epidemic} subset is able to act as additional benchmarks and contribute substantially to various research topics, such as pattern recognition and trend analysis.

\begin{figure*}[t]
\centering
\begin{subfigure}{\textwidth}
    \centering
    \includegraphics[width=.88\textwidth]{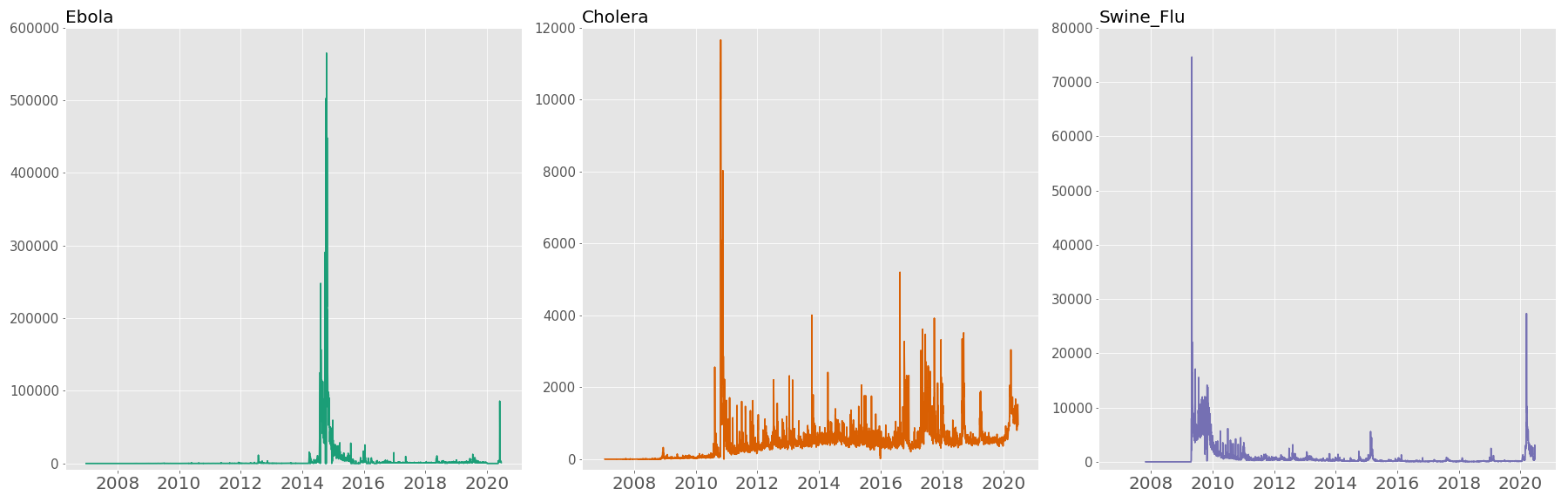}
    \caption{}
    \label{fig:line-general}
    \vspace{-.9mm}
    \includegraphics[width=.88\textwidth]{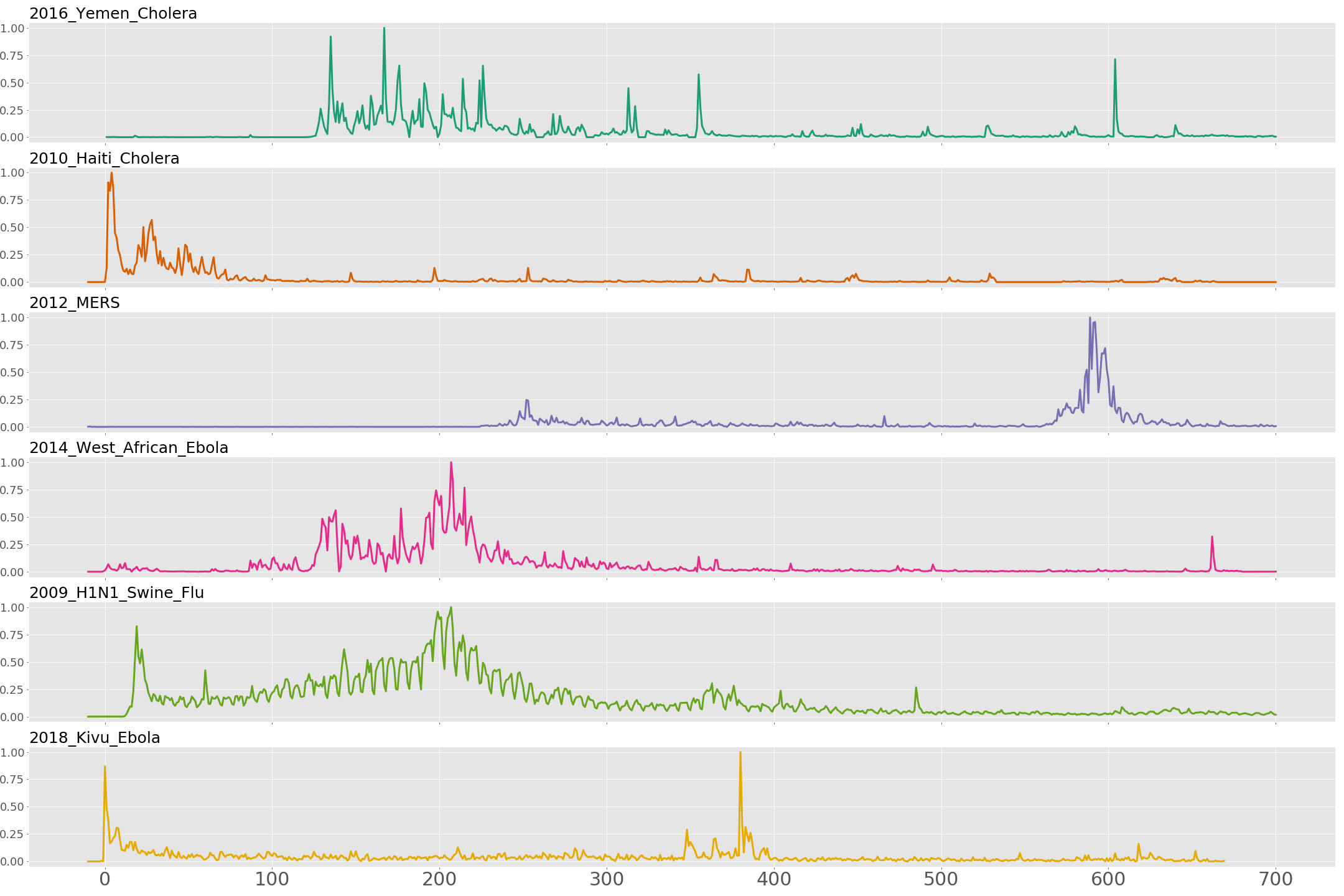}
    \caption{}
    \label{fig:line-outbreak}
    \vspace{-1.1mm}
\end{subfigure}
\caption{Time-series trend analysis. (a) Each chart represents a class of a general disease, where the x-axis represents the time as yearly dates, and the y-axis represents the corresponding number of tweets. (b) Each chart represents an outbreak class in the corpus. The x-axis represents a range of dates from day -10 to 700, where day 0 represents the respective start date of each outbreak, such as: 2016-09-28 (Yemen Cholera), 2010-10-20 (Haiti Cholera), 2012-09-23 (MERS), 2014-03-23 (West African Ebola), 2009-04-12 (Swine Flu), and 2018-08-01 (Kivu Ebola)\protect\footnotemark. The y-axis represents the number of tweets, normalized to between 0 and 1. }
\label{fig:lineplots}
\end{figure*}


\section{Data Exploration}

\footnotetext{According to the World Health Organization, https://www.who.int}

\subsection{Hashtags Analysis}

To gain a general overview of EPIC30M, we first conduct hashtags analysis for each epidemic and plot them on a 3 by 3 grid, as shown in Figure~\ref{fig:worldclouds}. The first row (Fig.~\ref{fig:wc-general}) represents three general diseases whereas the second and third rows (Fig.~\ref{fig:wc-outbreak}) represent the six outbreak classes in chronological order. Each word cloud contains the top 100 hashtags in their respective class, where the sizes represent their frequencies.

Through observation, we identify several interesting phenomena, such as: (1) Key terms provide semantic indication of the crises, in addition to possible cross-epidemic indicators: such as pandemic, epidemic, healthcare, vaccine, disease, sanitation, and others; (2) location-related hashtags, such as \textit{\#Yemen}, \textit{\#Haiti} and \textit{\#SierraLeone}, appear in all classes and occupy majority of the key words, which we believe to be the highest concerned feature; (3) several classes include hashtags of other diseases, i.e., \textit{\#COVID19} in the \textit{2016\_Yemen\_Cholera} class and \textit{\#Malaria} in the \textit{Cholera} class, which implies that discussions on cross-epidemic matters are popular; and (4) some hashtags refer to non-epidemic related events, such as \textit{\#5YearsOfWarOnYemen} and \textit{\#earthquake} appearing in the \textit{2016\_Yemen\_Cholera} and \textit{2010\_Haiti\_Cholera} sets respectively.

\subsection{Time-series Trend Analysis}

Subsequently, we conduct trend analysis with an attempt to identify time-variant patterns from the corpus. For the three general classes (Fig.~\ref{fig:line-general}), we plot each class into a line chart, where the x-axis represents the time in yearly dates and the y-axis represents the corresponding number of tweets. For the six outbreak classes (Fig.~\ref{fig:line-outbreak}), the x-axis of each line chart uses the number of days offset from the start date of the outbreak, whereas the y-axis represents the number of tweets normalized to between 0 and 1. 

Through the time-series line plots, we observe that some of the epidemics, i.e. \textit{2010 Haiti Cholera} and \textit{2018 Kivu Ebola}, show a surge in tweets before the respective official start dates of the outbreaks, which signifies the importance of leveraging social media to conduct early signal detection. We also observe that an epidemic outbreak not only leads to rapid discussion of its own, but also trigger exchanges of other diseases. Finally, the time-series analyses also show clear dynamic properties or trends with exponential increases (shocks or spikes) in tweet type and a temporal persistence after an initial shock~\cite{hamilton1994time}. Other dynamic properties that may be of interest include local cycles and trends.  Such dynamic effects, when paired with semantic content (such as healthcare related terms), may provide potential indicators of an onset of a crisis.

\section{Potential Use Cases}

While Twitter has an enormous volume and frequency of information exchange, i.e. over half a billion tweets posted daily, such rich data potentially exposes information on epidemic events through substantial analysis. In this section, we demonstrate the value and impact that EPIC30M could create by discussing on multiple use cases of cross-epidemic research topics that attract growing interests in recent years. These use cases span multiple research areas, such as epidemiological modeling, pattern recognition, natural language processing and economical modeling. We claim that EPIC30M fills the gap in the literature where very little disease related corpora are sizable and rich enough to support such cross-epidemic analysis tasks. EPIC30M supplies benchmarks of multiple epidemics to facilitate a wide range of cross-epidemic research topics. 

\textbf{Epidemiological Modeling}.
Epidemiological modeling provides various potential applications to understand the Twitter dynamics during and post-outbreaks, such as compartmental modeling~\cite{anderson2013compartmental} and misinformation detection~\cite{wu2016mining}. To name a few, \citet{jin2013epidemiological} uses Twitter data to detect false rumors and a susceptible-exposed-infected-skeptic (SEIZ) model to group users in four compartments. \citet{skaza2017modeling} use susceptible-infectious-recovered (SIR) epidemic models on Twitter' hashtags to compute infectiousness of a trending topic. In the recent event of COVID-19, these models are repeatedly applied to predict discrete questions, such as ~\citet{chen2020covid}'s proposal of using a time-dependent SIR model to estimate the total number of infected persons and the outcomes, i.e., recovery or death. 

\textbf{Trend Analysis and Pattern Recognition}.
Extensive prior works leverage social media data to perform trend analysis and pattern recognition tasks. For instance, \citet{kostkova2014swineflu} study the 2009 swine-flu outbreak and demonstrates the potential of Twitter to act as an early warning system up-to a period of two or three weeks. Similarly, \citet{joshi2020automated} predict alerts of Western Africa Ebola epidemic, three months earlier than the official announcement. While early detection and warning systems for crisis events may reduce overall damage and negative impacts~\cite{liu2020crisisbert}, EPIC30M provides high volume and timely information that facilitate trend analysis and pattern recognition tasks for epidemic events.

\textbf{Sentiment and Opinion Mining}.
The observation of social sentiments and public opinions plays an important part in benchmarking the effect of releasing public policy amendments or new initiatives. Several prior works leverage sentimental analysis and opinion mining to extract the contextual meaning of social media content. For instance, \citet{beigi2016overview} provides an overview of the relationship among social media, disaster relief and situational awareness in crisis time, and \citet{neppalli2017sentiment} performs location-based sentimental analysis on tweets for Hurricane Sandy in 2012.

\textbf{Topic Detection}.
Topic detection or modeling may enable authorities in anticipating a crisis and taking actions during the same. The technique helps in recognizing hidden patterns, understanding semantic and syntactic relations, annotating, analyzing, organizing, and summarizing the huge collections of textual information. Considering the same, several researchers have implemented these approaches on crises datasets to detect and categorize the potential topics. \citet{chen2016syndromic} suggest two topic modeling prototypes to ameliorate trends estimation by seizing the underlying states of a user from a sequence of tweets and aggregating them in a geographical area. In~\citep{lim2017clustop} researchers perform optimized topic modeling using community detection methods on three crises datasets~\cite{olteanu2014crisislex, olteanu2015expect, zubiaga2016analysing} to identify the discussion topics. 

\textbf{Natural Language Processing}.
Several works leverage Twitter datasets to conduct Natural Language Processing (NLP) tasks. As a challenging downstream task of NLP, Automatic Text Summarization techniques extract latent information from text documents where the models generates a brief, precise, and coherent summary from lengthy documents. Text summarization is applicable in various real-would activities during crisis, such as generating news headlines, delivering compact instructions for rescue operations and identifying affected locations. Prior works demonstrate such applications during crisis time. For instance, \citet{rudra2015extracting} and ~\cite{rudra2016summarizing} propose two relevant methods that classify and summarize tweets fragments to derive situational information. More recently, \citet{sharma2019going} proposes a system that produces highly accurate summaries from the Twitter content during man-made disasters. Several other works focus on NLP subtasks of social media, such as information retrieval~\cite{imran2014aidr,liu2020self} and text classification~\cite{parilla2014automatic, liu2019ipod}. 

\textbf{Disease Classification}.
Applications of Machine Learning and Deep Learning in the healthcare sector gather growing interests in recent years. For instance, ~\citet{krieck2011new} analyzes the relevance of Twitter content for disease surveillance and activities tracking, which help alert health official regarding public health threats. \citet{lee2013real} conducts text mining on Twitter data and deploys a real-time disease tracking system for flu and cancer using spatial, temporal information. \citet{ashok2019machine} develops a disease surveillance system to cluster and visualise  disease-related tweets.

\textbf{Crisis-time Economic Modeling}.
Estimating economical impact of crises, such as epidemic outbreaks, is a crucial task for policy makers and business leaders to adjust operational strategies~\cite{liu2019strategic} and make right decisions for their organizations in the time of crises. Several research studies in such domain. For instance, \citet{okuyama2008critical} provides an overview and a critical analysis of the methodologies used for estimating the economic impact of disaster; \citet{avelino2019challenge} proposes the Generalized Dynamic Input-Output framework (GDIO) to dynamically model higher-order economic impacts of disruptive events. Such studies correlate disaster events and economy impact, which rely on disaster-related data and financial market data, respectively. We believe that EPIC30M is able to contribute to future economic modeling studies for epidemic events.  

\textbf{Health Informatics}.
Compared to the cases above, a more general use case area is healthcare Informatics , i.e., “the integration of healthcare sciences, computer science, information science, and cognitive science to assist in the management of healthcare information”~\cite{mccormick2015essentials, avinash2007system, siau2006mobile}.  While social media and online sources are used to connect with patients and provide reliable educational content in health informatics, there is growing interest in using Twitter and other feeds to study and understand indicators for health trends or particular behaviors or diseases.  For example, \citet{nambisan2015social} utilize Twitter content to study the behavior of depression.  EPIC30M contains behavioral information across various diseases and how the populace behaves with the onset and persistence of the diseases.  Multiple disease cases will provide such research to correlate behavioral information across instances.

\textbf{News and Fake News}.  With the proliferation of news content through internet and virtual media, there is a growing interest in developing an understanding of the science of news and fake news~\cite{lazer2018science}.  Data mining algorithms are advancing to study news content~\cite{shu2017fake}.  EPIC30M contains real news content that grows over time from social lay-person terminology to technical and professionally based information and opinion. It likewise includes fact-based information as well as distorted or fake content.  Through multiple cases over time, the field will have a rich source to study news content, especially when correlating with reliable news sources for particular snapshots of time.

All in all, we believe that EPIC30M provides a set of rich benchmarks and is able to facilitate extensions of the above-mentioned works on a higher order, e.g., in cross-epidemic settings.  As a result, the research findings are more robust and closer to real-world scenarios.

\section{Conclusion and Future Work}

\textbf{Conclusion}. During our other efforts on COVID-19 related work, we discovered very little disease related corpora in the literature that are sizable and rich enough to support such cross-epidemic analysis tasks. In this paper, we present EPIC30M, a large-scale epidemic corpus that contains 11.8 millions tweets from 2006 to 2020. The corpus includes a subset of tweets related to three (3) general diseases and another subset related to six (6) epidemic outbreaks. We conduct exploratory analysis to study the properties of the corpus and identify several phenomena, such as strong correlation between epidemics and locations, frequent cross-epidemic topics, and surge of discussion before occurrence of the outbreaks. Finally, we discuss a wide range of use cases that EPIC30M can potentially facilitate. We anticipate that EPIC30M brings substantial value and impact to both fast growing computer science communities, such as natural language processing, data science and computation social science, and multi-disciplinary areas, such as economic modeling, health informatics and the science of news and fake news.

\textbf{Future work}. For some epidemic outbreaks, such as \textit{2009 H1N1 Swine Flu} and \textit{2014 West Africa Ebola}, EPIC30M includes relevant tweets posted throughout the respective duration of the epidemics. We expect the data of these few classes could serve as strong and timeless cross-epidemic and cross-disease benchmarks. On the other hand, several epidemics, such as \textit{2018 Kivu Ebola} and \textit{2016 Yemen Cholera}, are still ongoing. We intend to extend the corpus by actively or periodically crawling tweets in addition to the current version. Furthermore, we plan to further develop the corpus with additional epidemic outbreak classes that happened more recently, such as the \textit{2019 multi-national Measles outbreaks} in the DR Congo, New Zealand, Philippines and Malaysia, the \textit{2019 Dengue fever epidemic} in Asia-Pacific and Latin America, and the \textit{2018 Kerala Nipah virus outbreak}. Lastly, we also intend to develop an active crawling web service that automatically update EPIC30M, and migrate to cloud-based relational database services to ensure its availability and accessibility.

The corpus is available at https://www.github.com/junhua/epic.


\section{Acknowledgement}

This research is funded in part by the Singapore University of Technology and Design under grant SRG-ISTD-2018-140.

\bibliographystyle{ACM-Reference-Format}
\bibliography{ref}

\end{document}